\documentclass[final 
    ,final            
  ]
  {aipproc}
\usepackage{amsmath, amsfonts, amssymb,latexsym }

\layoutstyle{8x11single}

\newcommand{\B}{{B^{(1)}}}

\begin{document}

\title{Gamma-ray signatures for Kaluza-Klein dark matter}


\author{Lars Bergstr\"om}{
  address={Department of Physics, AlbaNova University Center, Stockholm
       University, SE - 106 91 Stockholm, Sweden}
}

\author{Torsten Bringmann\footnote{email: \texttt{bringman@sissa.it}} $^\textrm{\footnotesize ,}$}{
  address={SISSA/ISAS, via Beirut 2-4, 34013 Trieste, Italy}
  ,altaddress={Department of Physics, AlbaNova University Center, Stockholm
       University, SE - 106 91 Stockholm, Sweden}
}

\author{Martin Eriksson}{
  address={Department of Physics, AlbaNova University Center, Stockholm
       University, SE - 106 91 Stockholm, Sweden}
}

\author{Michael Gustafsson}{
  address={Department of Physics, AlbaNova University Center, Stockholm
       University, SE - 106 91 Stockholm, Sweden}
}

\begin{abstract}
  The extra-dimensional origin of dark matter is a fascinating and nowadays often discussed possibility. Here, we present the gamma-ray signatures that are expected from the self-annihilation of Kaluza-Klein dark matter particles. For comparison, we contrast this with the case of supersymmetry, where the neutralino annihilation spectra take a very different form. In both cases we find pronounced spectral signatures that could in principle be used to distinguish between these two types of dark matter candidates already with today's detector resolutions.
\end{abstract}

\maketitle


\section{Introduction}

While the existence of dark matter is by now widely accepted, with evidence coming from various independent observations over an impressive range of distance scales, its nature still remains unknown, providing a major challenge for cosmology and astroparticle physics. One of the most favourate candidates is certainly the supersymmetric neutralino, theoretically well motivated by considerations from particle physics (for reviews, see, e.g. \cite{chirev}).

More recently, the proposal of an alternative dark matter candidate \cite{CMS,STa} has received much attention. It arises naturally in models with universal extra dimensions (UED) \cite{app}, where all standard model fields are allowed to propagate in a higher-dimensional bulk. After compactification of the internal space, these additional degrees of freedom appear as towers of new, heavy states in the effective four-dimensional theory. The lightest of these Kaluza-Klein particles (LKP) is usually stable due to KK parity, an internal ${Z}_2$ symmetry that appears as a remnant of higher-dimensional translational invariance. Taking into account radiative corrections to the KK masses, the LKP is expected to be well approximated by the $\B$, the first KK excitation of the weak hypercharge boson \cite{CMS}. Detailled relic density calculations \cite{KM} show that it can account for the required dark matter density as determined by WMAP \cite{wmap} if the compactification scale (and thus the $\B$ mass) lies in the TeV range, the precise value depending on the exact mass spectrum and the resulting coannihilation channels. This is particularly interesting since it means that the LHC, which will probe compactification scales up to about 3 TeV \cite{LHCbound}, will have access to practically the whole cosmologically relevant parameter space. The prospects for both direct and indirect LKP detection have been studied in some detail and generally found to be quite promising, at least for next generation's detectors \cite{oldfrag,LKPdet,BBEGa,BBEGb}.

The underlying idea of indirect detection is always to look for exotic contributions to cosmic rays that might originate from  the  self-annihilation of dark matter particles. Unfortunately, total fluxes are notoriously difficult to predict due to the largely unknown distribution of dark matter. For gamma rays, in particular, these astrophysical uncertainties result in an overall \emph{amplitude} that may vary over several orders of magnitude \cite{FPS}. The \emph{spectral form}, on the other hand, is unaffected and can be predicted with great accuracy from the underlying particle physics. It is thus of great interest to extract pronounced spectral signatures that could unambiguously identify dark matter contributions to astrophysical gamma-ray signals and maybe even allow to discriminate between different dark matter candidates. Here, we present two such signatures, a sharp cutoff at the dark matter particle's mass and the line signal from direct annihilation into gamma rays, and compare the situation for the LKP and the neutralino. With a massive vector versus a Majorana fermion particle, note that this actually entails a comparison between two different \emph{types} of dark matter particles so that the results should not be restricted to the specific models studied here.

\section{A sharp cutoff from final state radiation}

\begin{figure}[t]
\includegraphics[width=0.5\columnwidth]{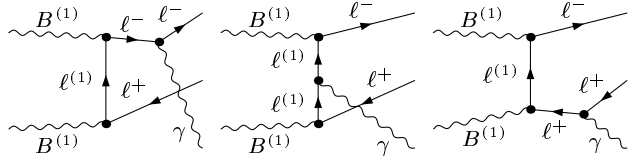}
\caption{Contributions to $\B\B \rightarrow
\ell^+ \ell^- \gamma$. In principle, there are also annihilation diagrams with a second KK-level Higgs in the $s$-channel, but these are usually heavily suppressed by a factor $m_\ell^2/(s-M_{H^{(2)}}^2) $ unless the $H^{(2)}$ mass is fine-tuned to be near the resonance.
}
\label{fig_LKPfsr}
\end{figure}

The first example to be considered here is that of final state radiation (FSR for short, sometimes also dubbed \emph{internal bremsstrahlung}), where an additional photon is emitted in situations when a pair of dark matter particles decays into a pair of charged standard model particles. The $\B$, in particular, annihilates mainly  into quarks ($\sim$35 \%) and charged leptons ($\sim$59 \%), giving rise to the FSR diagrams shown in Fig.~\ref{fig_LKPfsr}. The spectral distribution of the photons produced in this way is well approximated by \cite{BBEGa}
\begin{equation}
  \label{FSRf}
  \frac{\mathrm{d}N_\gamma^\ell}{\mathrm{d}x} \equiv
  \frac{\mathrm{d}(\sigma_{\ell^+\ell^-\gamma})/\mathrm{d}x}{\sigma_{\ell^+\ell^-}}
  \simeq \frac{\alpha}{\pi} \frac{(x^2 - 2x + 2)}{x} \ln{\left[
  \frac{m^2_{\B}}{m^2_\ell}(1-x) \right]},
\end{equation}
where $x\equiv E_\gamma / m_{\B}$. In fact, this is the universally expected spectral form for FSR in the case of final state fermions and center-of-mass energies much larger than the mass of the final state particles \cite{BBEGa,bir}; in this case one expects a factor $\alpha/\pi$ from the electromagnetic coupling
and the phase space difference between two- and three-body final
states, multiplying a collinear term containing a large logarithm. From the form of (\ref{FSRf}) it is evident that light leptons ($e^+e^-$ pairs, in particular) give the dominant contribution to FSR photons in the LKP scenario.

\begin{figure}
\includegraphics[width=0.6\columnwidth]{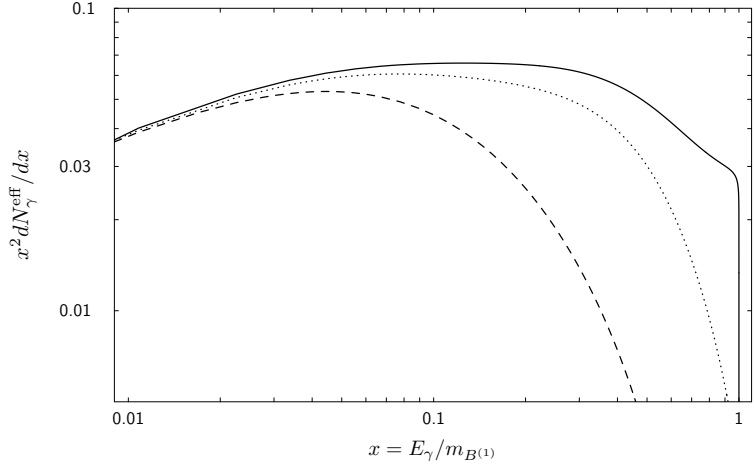}
\caption{The total number of photons per $\B\B$ annihilation (solid
line), multiplied by $x^2=(E_\gamma/m_{\B})^2$. Previously, one has only considered the contribution from quark fragmentation alone (dashed line, see also \cite{oldfrag}). The dotted line shows the contribution from both quark fragmention and the semihadronic decay of final state $\tau$ leptons. At the highest energies, the total spectrum is dominated by photons from internal bremsstrahlung off light leptons. (Figure taken from \cite{BBEGa}).}
\label{spectrum}
\end{figure}

Traditionally, one has neglected photons from leptonic final states and only considered secondary photons from quark fragmentation and decay, where the main contribution comes from neutral pion decay \cite{oldfrag}. The result is a featureless, rather soft  photon spectrum. FSR photons from light leptons, on the other hand, show a much harder spectrum, with a sharp cutoff at the $\B$ mass. 
In Fig.~\ref{spectrum}, we compare the relative importance of these processes for the total annihilation spectrum (for the fragmentation spectra we use parametrizations \cite{param} based on \textsc{Pythia} \cite{PYTHIA} runs; note that the effect of FSR from both quarks and further decay products is already included here). We find that for the highest energies  FSR photons from light leptons clearly dominate the spectrum -- a result that can be traced back to the large branching ratio into these states as well as the high mass of the $\B$, leading to a large logarithmic enhancement in (\ref{FSRf}).

A rather hard spectrum combined with a sharp cutoff like this provides a feature that is hard to mimic with astrophysical processes and would thus provide a smoking gun signal for the annihilation of dark matter particles. Originally it had in fact been proposed that the TeV gamma-ray signal from the galactic center, as observed by the HESS telescope \cite{hess}, could be explained in terms of a similar (though heavier) dark matter candidate with the same annihilation spectrum as the $\B$ \cite{BBEGa}. The new HESS data, however, show a spectrum that is well described by a power law with spectral index 2.3 in the region  from 100 GeV to at least 20 TeV \cite{hessdata} -- which makes an astrophysical explanation much more likely and disfavours any attempt to attribute a dark matter related origin to the whole range of observed gamma rays. It is nevertheless interesting to note that the expected gamma-ray flux from  $\B$ annihilations turns out to be of the right order of magnitude to potentially give visible distortions in the background flux; with extended observation times and energy resolutions, it is thus in particular worthwhile to watch out for edge-like features in the HESS signal that could be related to the sharp cutoff shown in Fig.~\ref{spectrum}.

\section{The line signal}

Dark matter has to be electrically neutral, so the direct (pair-) annihilation into $\gamma\gamma$, $\gamma Z$ or $\gamma H$ final states is always loop-suppressed. Since the annihilating particles are highly non-relativistic, such a process would on the other hand result in a very sharp annihilation line, providing a spectacular gamma-ray signature that, if detected, could hardly be mistaken for anything else.  

\begin{figure}[t]
\includegraphics[width=0.6\textwidth]{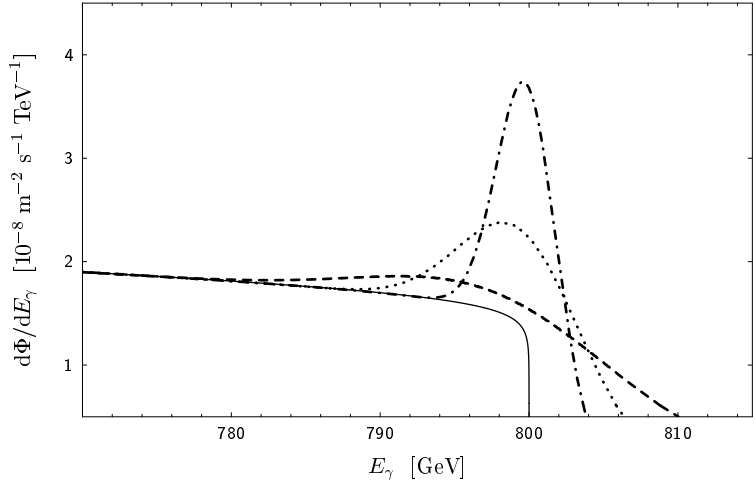}
\caption{The line signal from $\B\B\rightarrow\gamma\gamma$ annihilations in the galactic center, as it would be seen by a detector with an energy
resolution of $2\sigma$ = 2 \% (dashed), 1 \% (dotted) and 0.5 \% (dash-dotted), respectively. The expected flux from final state radiation is plotted as a solid line. See the text for more details. (Figure taken from \cite{BBEGb}).}
\label{peakobs}
\end{figure}

The line signal from LKP dark matter has been studied in \cite{BBEGb}, where a full one-loop calculation was performed. Unlike in the situation for FSR, the cross section here depends on the details of the annihilation process; in particular, it is rather sensitive to the mass splittings between the LKP and the charged Kaluza-Klein particles that appear in the loop. 
As an example, we plot in Fig.~\ref{peakobs} the expected flux from $\B\B\rightarrow\gamma\gamma$ annihilations in the galacic center, with $m_{\B}=800$ GeV and a mass splitting of $m_{f^{(1)}}/m_{\B}=1.05$, together with the FSR contribution that was discussed in the previous section. (In this particular example, we have used a dark matter profile that takes into account the expected effects of baryonic compression \cite{prada};  the \emph{ratio} of the plotted fluxes is of course not affected by any astrophysical assumptions about the dark matter distribution).
We conclude that, unfortunately, in order to be able to actually resolve the line feature, one would need much better energy resolutions than the 15\% that are typical for air Cerenkov telescopes (ACTs) like HESS.\footnote{
Smaller mass splittings result in an enhancement of the line signal. The value of 5\% that we used here is actually rather conservative since the amplitude is dominated by loops with right-handed KK leptons, which in the minimal scheme of calculating radiative corrections to KK masses are expected to be only about 1\% heavier than the $\B$ \cite{CMS}. Leaving the mimimal UED scheme, even smaller mass splittings can be considered \cite{KM}. For realistic values, however, the maximally expected enhancement of the annihilation line does not exceed a factor of about two. 
}

The direct annihilation of a $\B$ pair into $\gamma Z$ or $\gamma H$ is also possible. Since $m_\B\ll m_{Z,H}$, the resulting gamma-ray lines can not be resolved separately, but simply add to the $\gamma\gamma$ contribution shown in Fig.~\ref{peakobs}. The diagrams contributing to the $\gamma Z$ process have a very similar 
structure as for the $\gamma\gamma$ case -- except
for the fact that $Z$ bosons also have an \emph{axial} vector part in
their coupling to fermions. Taking this into account, one can compare
the $\gamma$ and $Z$ coupling strengths to obtain as a quick semi-analytical estimate that
the $Z\gamma$ process should enhance the gamma ray signal shown in
Fig.~\ref{peakobs} by about 10\%; we have confirmed this estimate 
numerically. Finally, the $H\gamma$ line is interesting since it is absent for neutralinos and appears as a new feature of Kaluza-Klein dark matter. Unfortunately, the number of contributing diagrams with non-supressed couplings is much smaller than in the case of the $\gamma\gamma$ or $\gamma Z$ lines, so one should expect its contribution to the total strength of the line signal to be subdominant.

\section{A comparison with neutralino dark matter}

Let us now compare the spectral features for LKP dark matter with those of neutralino dark matter.
While the $\B$ is a massive vector particle, the neutralino is a Majorana fermion. As an important consequence, its annihilation into light leptons is helicity-suppressed so that in this case we cannot expect a large contribution of the form (\ref{FSRf}) to the gamma-ray spectrum. Depending on the parameters of the supersymmetric model, however, a considerable branching ratio can go into charged gauge bosons, making it worthwhile to investigate in detail the resulting FSR spectrum also for this case.

We will focus in the following on rather heavy (TeV scale) neutralinos, which is motivated by three independent reasons. The first one is related to the fact that there is a considerably sized region in the parameter space of the minimally supersymmetric standard model (MSSM) where the lightest supersymmetric particle is a TeV scale neutralino with the right relic density to provide all the dark matter.\footnote{
In the hyperbolic branch/focus-point region of minimal supergravity (mSUGRA), for example, neutralino masses extend up to at least $m_\chi\sim1.2$ TeV \cite{BKPU};  in the minimal anomaly mediated supersymmetry breaking (mAMSB) scheme, one expect masses slightly above 2 TeV \cite{MPU}. Leaving minimal models for supersymmetry breaking, but still staying within the MSSM, one can easily find even higher neutralino masses \cite{pro}.
}
 In these situations, however, supersymmetry would most likely be outside the kinematical reach of the LHC. Also direct detection rates fall off rapidly with higher masses, so it is particularly important to look for promising signatures that would allow a possible indirect detection. Secondly, we want to compare the case of neutralinos with that of LKP dark matter. In particular, we want to investigate whether there is a signifant difference in the gamma-ray spectra near the cutoff at the dark matter particle's mass, which is expected to be at the TeV scale for the LKP. Finally, just as in the case of lepton final states, we expect the contribution from FSR to be more significant for higher center-of-mass energies; also, a heavy neutralino usually has a large branching ratio into charged gauge bosons.

\begin{figure}[t]
  \includegraphics[angle=270,width=0.5\textwidth]{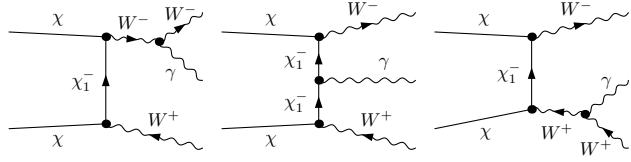}\\
  \caption{Contributions to $\chi\chi \rightarrow W^+ W^- \gamma$ for a pure higgsino-like neutralino (crossing fermion lines are not shown).}
  \label{feyn_SUSY}
\end{figure}

The neutralino is given by a linear combination of the superpartners of the gauge and Higgs fields,
\begin{equation}
  \chi\equiv\tilde\chi^0_1= N_{11}\tilde B+N_{12}\tilde W^3 +N_{13}\tilde H_1^0+N_{14}\tilde H_2^0\,.
\end{equation}
In order not to overclose the universe, however, a thermally produced TeV-scale neutralino will generally not be a mixed state but an almost pure Higgsino (when the usual GUT condition $M_1\sim M_2/2$ is imposed, like in the case of mSUGRA) or Wino (e.g.~in the case of mAMSB). In the pure Higgsino or Wino limits, the analytical expression for FSR from charged gauge boson final states simplifies considerably since all couplings take rather simple forms and only a limited number of Feynman diagrams contributes to the process $\chi\chi \rightarrow W^+ W^- \gamma$, as shown in Fig. \ref{feyn_SUSY} for the case of a Higgsino.

For vector boson final states, the FSR spectrum cannot easily be cast into a universal form as for fermionic final states \cite{bir}, but instead depends on the internal structure of the annihilation process. If the $W^\pm$ bosons are radiated from a fermion line as in the situation depicted in Fig.~\ref{feyn_SUSY}, an interesting effect leads to an enhanced production of high-energy FSR photons as a consequence of the IR behaviour of QED: for kinematical reasons, a high-energy photon is automatically accomanied by a low-energy $W$ boson, which can be treated like a ``soft photon'' in the limit of high neutralino masses. For a more detailled discussion, including analytical expressions for the spectrum, see \cite{BBEGc}.

Besides FSR, there are more traditional sources for gamma rays from neutralino annihilations. As discussed before, the dominant contribution at energies considerably below the cutoff comes from fragmentation of the decay products and subsequent $\pi^0$ decay. Heavy neutralinos mainly annihilate into vector bosons and heavy quarks; we checked that in this case the resulting gamma-ray spectrum from fragmentation is almost independent of the exact branching ratios. At the neutralino mass, finally, one expects a line signal from the direct annihilation into $\gamma\gamma$ \cite{gg} and $Z\gamma$ \cite{Zg}, providing a spectacular signature in sharp contrast to the featureless fragmentation spectrum. In the high mass, pure higgsino or wino limit which we are interested in here, this line signal is enhanced owing to nonperturbative, binding energy effects \cite{his}, thus leading to promising observational prospects.

\begin{table}[t]
   \begin{tabular}{ccccccccccc}
   \hline
    $M_2$   &  $\mu$  &  $m_A$  &  $m_{\tilde f}$  &  $A_{f}$   &  $\tan \beta$  \;&
$m_\chi$ & $m_{\chi^\pm_1}$ & $Z_h$ & $W^\pm$ & $\Omega_\chi h^2$ \\
    \hline
    3.2 & 1.5 & 3.2 & 3.2 & 0.0 & 10.0 \;& 1.50 & 1.51 & 0.92 & 0.39 & 0.12\\
    \hline
   \end{tabular}
\caption{\label{tab_susy} A choice of MSSM parameters and the
resulting neutralino mass $m_{\chi}$, chargino mass
$m_{\chi^\pm_1}$, higgsino fraction $Z_h$ and branching ratio into
$W$ pairs. This model fulfills all experimental constraints and gives the right relic density $\Omega_\chi h^2$.}

 \label{tabmod}
\end{table}

\begin{figure}
  \begin{minipage}[t]{0.49\textwidth}
      \centering
  \includegraphics[width=\columnwidth]{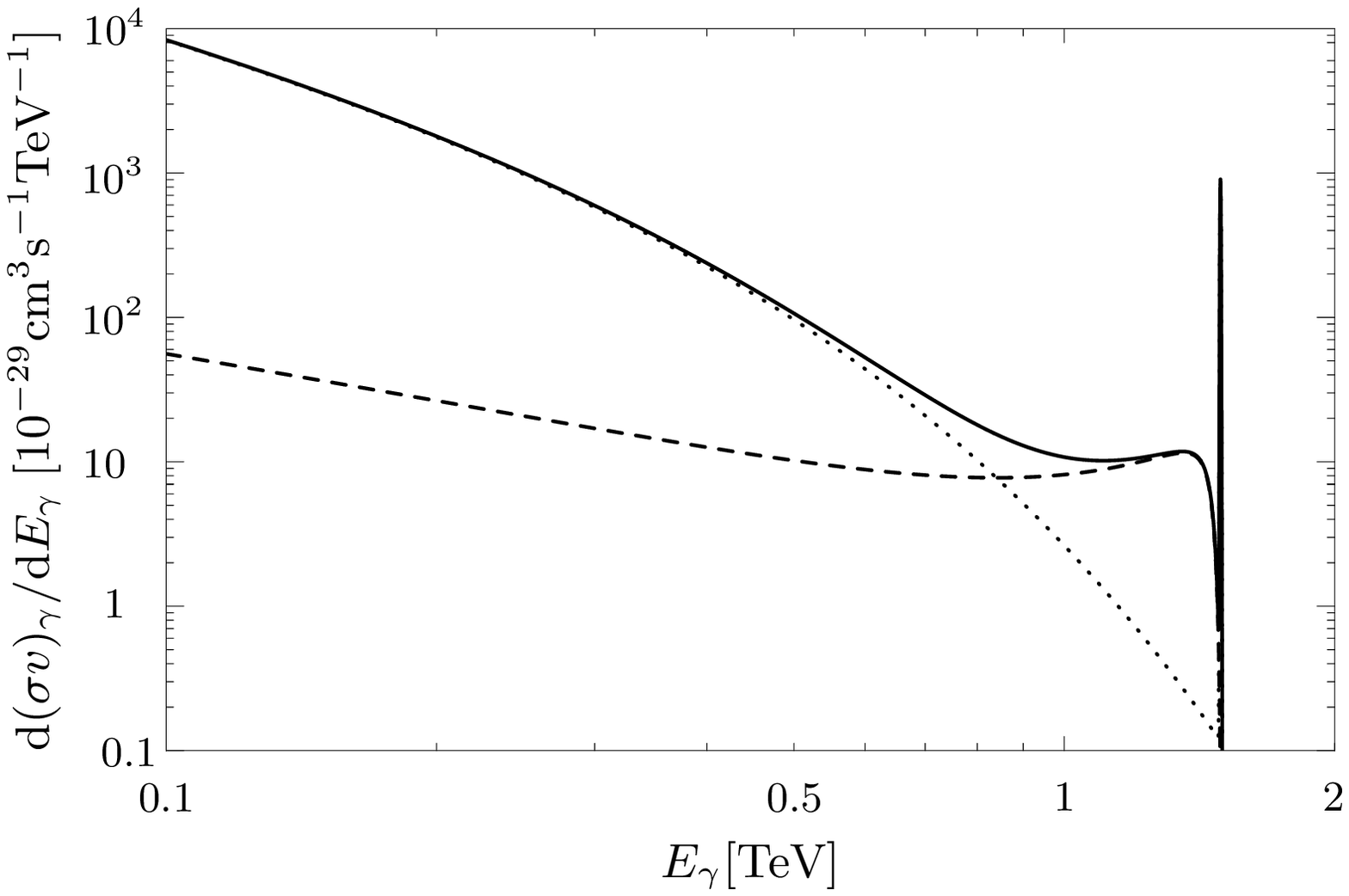}\\
   \end{minipage}
  \begin{minipage}[t]{0.02\textwidth}
   \end{minipage}
  \begin{minipage}[t]{0.49\textwidth}
      \centering   
  \includegraphics[width=\columnwidth]{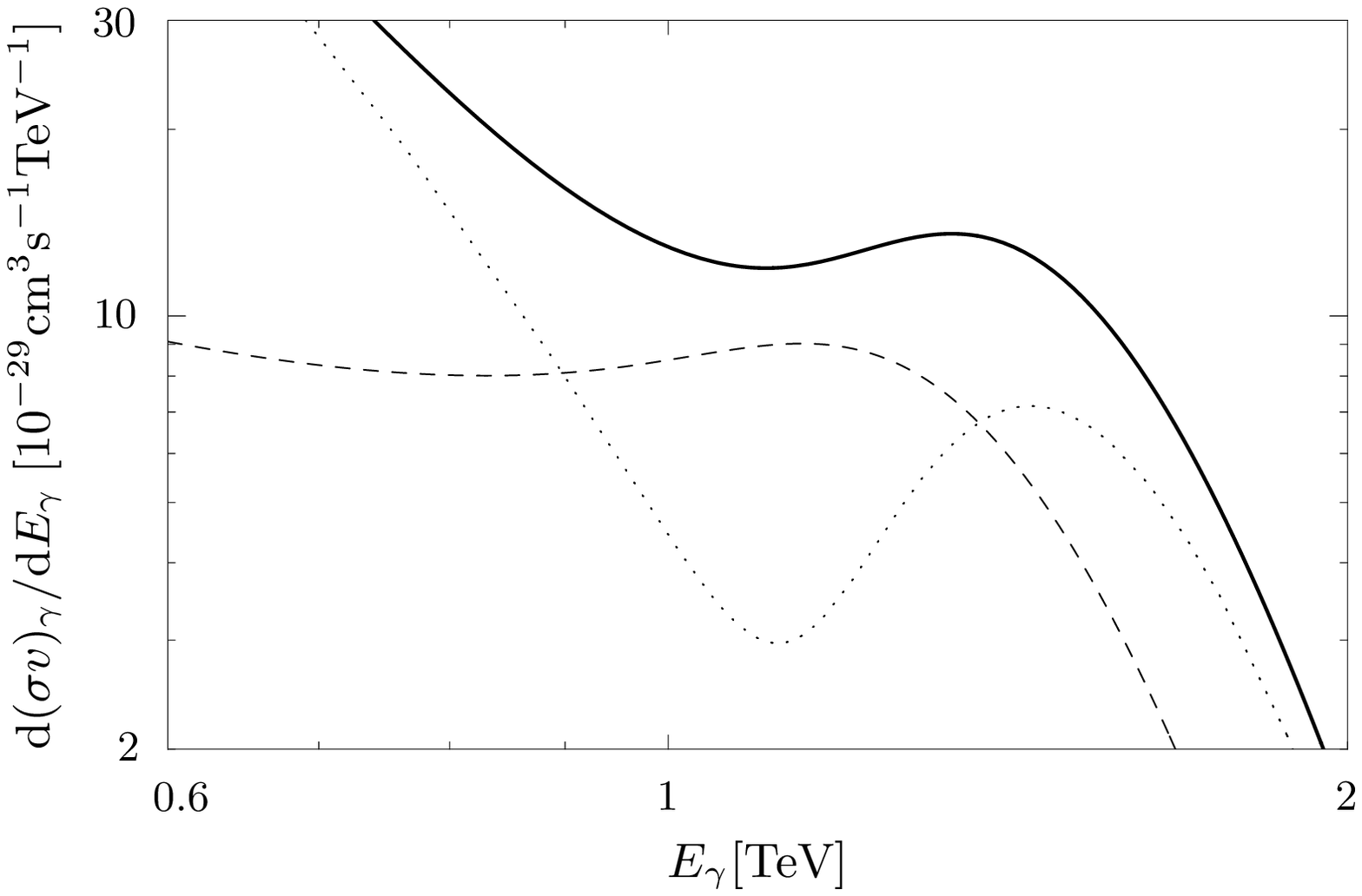}\\
   \end{minipage}
     \caption{To the left we plot the differential photon distribution for $\chi\chi$ annihilation (solid line), with MSSM model parameters as specified in Table \ref{tabmod}. Separately shown are the contributions from final state radiation (dashed), and  the combined contribution from the fragmentation of the decay products (mainly $W^+W^-$,  $ZZ$ and $b\bar b$) and the $\chi\chi\rightarrow\gamma\gamma, \, Z\gamma$ lines (dotted). The figure on the right shows the same situation, as seen by a detector with an energy resolution of 15 percent, which is typical for an ACT. (Figures taken from \cite{BBEGc}).}
     \label{susy_spec}
\end{figure}

In the following, we will consider the particular MSSM model specified in Table~\ref{tabmod}, chosen as an example that serves to illustrate the different contributions to the neutralino annihilation spectrum (very similar models can be found in the mSUGRA focus-point region). In Fig.~\ref{susy_spec}, we plot the total expected gamma-ray spectrum, taking into account fragmentation of the final states, FSR from $W^+W^-$ pairs and the line signal.\footnote{
 Note that the fragmentation of \emph{all} final states is taken into account -- which of course is also true for \cite{BBEGc}, even though the corresponding formulation used there might be a bit misleading in that respect.
 }
 Just as  in the case of the UED scenario, FSR photons clearly dominate over photons from fragmenation at the highest energies. Even with the energy resolution of today's detectors, and given that total rates are sufficient, one would be able to see their effect as both an enhancement of the annihilation line and a modification of the spectrum at slightly lower energies, making it a promising signature to look for. It is particularly interesting to compare the obvious difference in the annihilation spectra for LKP and neutralino dark matter, respectively.

\begin{figure}[t]
\includegraphics[width=0.6\textwidth]{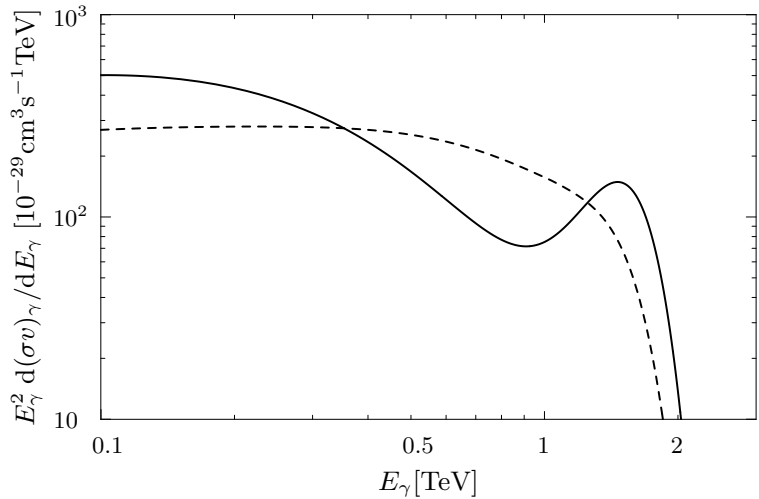}
\caption{The total gamma-ray spectrum that is expected from LKP (dashed line) and neutralino (solid line) dark matter annihilations, as seen by a detector with an energy resolution of 15 \% and including photons from fragmentation, FSR and the direct annihilation line. In order to facilitate the comparison, we have in both cases adopted a dark matter particle mass of 1.5 TeV and normalized the spectrum to a lowest order annihilation cross section of $\langle\sigma v\rangle_0=3\cdot 10^{-26}\mathrm{cm}^3\mathrm{s}^{-1}$.}
\label{compfig}
\end{figure}

\section{Conclusions}

Absolut dark matter annihilation fluxes are hard to predict due to the large astrophysical uncertainties involved. For a possible claim of an indirect detection of dark matter to be convincing, it should therefore be grounded on unambiguous spectral signatures.

We have presented such signatures for the gamma-ray spectrum from both Kaluza-Klein and supersymmetric dark matter: we find a sharp cutoff at the dark matter particle's mass, with features at slightly lower energies that would allow to distinguish between these two scenarios already with the energy resolution of typical ACTs in operation (this situation is visualized in Fig.~\ref{compfig}).
Since photons with the highest energies are dominantly produced by final state radiation, we have thereby also demonstrated the importance of taking into account this type of radiative corrections when considering dark matter annihilations -- a view that has been strongly advocated also in \cite{bir}.

The spectral features that we have reported here can be looked for wherever one expects enhanced dark matter densities, e.g.~in the direction of nearby dwarf galaxies or the galactic center \cite{FPS}; individual dark matter clumps or intermediate mass black holes \cite{imbh} in the galactic halo provide alternatives that are also well worth investigating. With the advance of new gamma-ray instruments of unprecedented size and energy resolution \cite{newgamma}, these will be exciting possibilities to further explore.


\end{document}